\documentclass{article}

\usepackage{amsmath}
\usepackage{microtype}
\usepackage{graphicx}
\usepackage{subfigure}
\usepackage{booktabs} 
\usepackage{amsfonts} 

\usepackage{hyperref}

\usepackage[accepted]{icml2022}

\icmltitlerunning{Submission and Formatting Instructions for ICML 2021}

\begin{document}

\twocolumn[
\icmltitle{Reduced Order Model for Chemical Kinetics: \\ A case study with Primordial Chemical Network}



\begin{icmlauthorlist}
\icmlauthor{Kwok Sun Tang}{astro}
\icmlauthor{Matthew Turk}{astro,ischool}
\end{icmlauthorlist}

\icmlaffiliation{astro}{Department of Astronomy, University of Illinois Urbana-Champaign, US}
\icmlaffiliation{ischool}{School of Information Sciences, University of Illinois at Urbana-Champaign, US}

\icmlcorrespondingauthor{Kwok Sun Tang}{kwoksun2@illinois.edu}

\icmlkeywords{Machine Learning, ICML}

\vskip 0.3in
]



\printAffiliationsAndNotice{}  

\begin{abstract}
Chemical kinetics plays an important role in governing the thermal evolution in reactive flows problems. The possible interactions between chemical species increase drastically with the number of species considered in the system. Various ways have been proposed before to simplify chemical networks with an aim to reduce the computational complexity of the chemical network. These techniques oftentimes require domain-knowledge experts to handcraftedly identify important reaction pathways and possible simplifications. Here, we propose a combination of autoencoder and neural ordinary differential equation to model the temporal evolution of chemical kinetics in a reduced subspace. We demonstrated that our model has achieved a close-to 10-fold speed-up compared to commonly used astro-chemistry solver for a 9-species primordial network, while maintaining 1 percent accuracy across a wide-range of density and temperature.
\end{abstract}

\section{Introduction}
\label{submission}

Chemistry plays a key role in regulating the cooling and the thermodynamical properties of the gas in astrophysical envirnoment. Chemical species are fundamental to our understanding of the formation of stars with different metalicities  \citep{Omukai_2000, Omukai_2005}, interstellar medium \cite{Gong_2017}, protoplanetary disk evolution \citep{Kamp_2017}, early universe \cite{2014MNRAS.440.3349R,2014MNRAS.442.2780R, Grackle} and etc.

The chemical network kinetics can in general be described as a initial value problem
\begin{align*}
    \frac{d \mathbf{y}}{dt} &= \mathbf{f}(t,\mathbf{y}), \\
    \mathbf{y}(t) &= \mathbf{y}(0) + \int_0^{t} f (t',\mathbf{y}(t')) dt',
\end{align*}
where $\mathbf{y}(t)$, the state vector, corresponds to the chemical abundance, and the thermal energy at a given instant in time $t$. $f$ specifies the interactions and dynamics among different species and guides the evolution of the state vector with time.


Scientific communities have been taking advantage of the recent advancement in deep learning \cite{champion2019, VinuesaBrunton2021, ml_fluid, Vlachas2022} to model dynamical systems. Many, if not all, of these approaches seek to identify the low-dimensional coordinates for an inherently high-dimensional systems that can reconstruct the dynamics in the physical space with less computational cost. It has a long history in the community of fluid dynamics. These are often termed as reduced-order models and are closely related to dimension reduction techniques like principal orthogonal decomposition, principal-component analysis, dynamic-model decomposition. The introduction of autoencoders \cite{BALDI198953, GoodBengCour16} generalized these powerful techniques from learning linear subspace to learning coordinates on a curved manifold and have shown to improve greatly the performance of classical ROM models (\citet{VinuesaBrunton2021} for a detailed recent review). 


Recently, data-driven methods have been introduced to study chemistry in astrophysical environments. \citet{grassi2021} applied the autoencoder with a combination of a latent fiducial chemical network to simplify a 29-species chemical network with 224 reactions into a reduced network with 5 species and 12 reactions. Their work has demonstrated a significant (65-times) speed-up. However, this study is limited to a constant density, temperature and cosmic-ionization rate environment.


In this work, we expand on the model proposed by \citet{grassi2021} with a modified architecture and  apply it to a 9-species primordial chemical network problem. Instead of focusing on a fixed density, and temperature grid, our proposed model is trained and evaluated on a snapshot taken from a full cosmological volume, with density ranging from $10^{-28} - 10^{-12} ~ \rm g ~ cm^{-3}$ and temperature ranging from $\rm 50 - 2000 K$. We address the two questions: 1.How accurate could neural network based reduced system be? 2. How much speed up could we gain with reduced models?

\section{Data}
To train our proposed model, physically realistic initial conditions are taken from a snapshot of a full cosmological simulations of first stars with the open-source simulation codebase \texttt{Enzo} \cite{Bryan_2014}. The density and temperature considered spans 14 and 2 orders of magnitudes. The detail of the simulations and the phase-space distribution of the initial conditions are outlined and shown in Appendix \ref{app:sims}.  The chemical abundances and thermal energy from the simulation grid cell are taken as initial conditions and are evolved for one freefall time with logarithmically-spaced timesteps to capture the wide range of dynamical behavior across different timescale with the meta-solver \texttt{Dengo} \cite{dengo} and the ODE integrator \texttt{Sundials CVODE} \cite{hindmarsh2005sundials}. Note that the spatial information of each cell is discarded. In this study, we have limited ourselves to studying the 9-species network which includes $\rm H_2, H_2^+, H, H^+, H^-, He, He^+, He^{++}, e^-$ and thermal energy.  The total state space vector $\mathbf{x} \in \mathbb{R}^{N}$ lives in 10 dimension space.
We refer interested readers to the \texttt{Grackle} method paper \cite{Grackle} for a more in-depth discussion of the 9-species network.  





\begin{figure}[!ht]
\begin{center}
\centerline{\includegraphics[width=\linewidth]{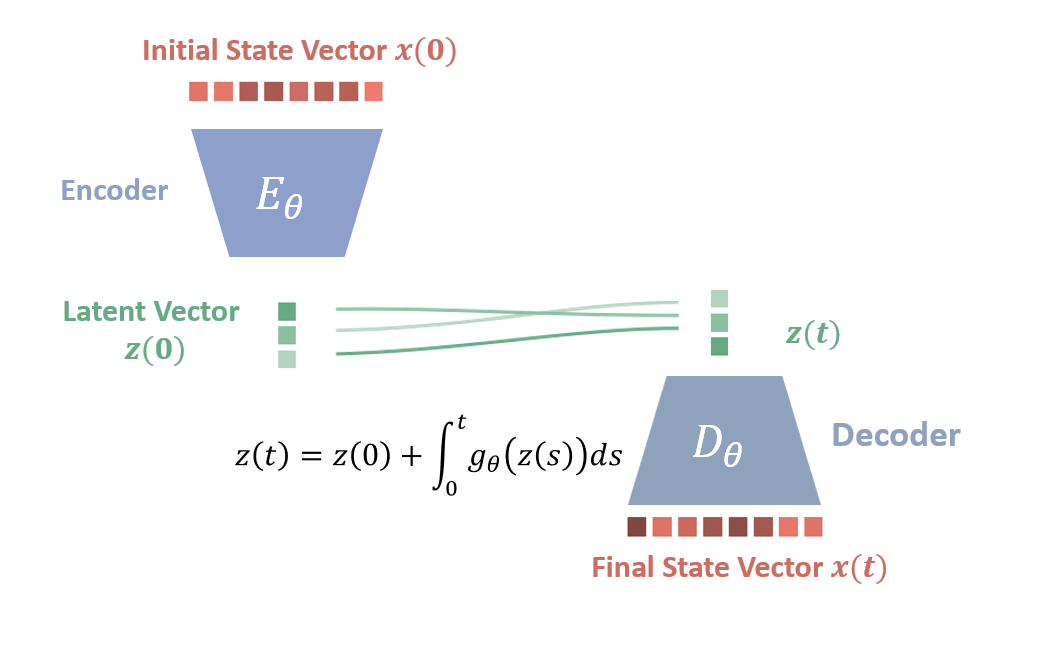}}
\vspace{-0.5cm}
\caption{A Schematic of our proposed model to evolve the system for one timestep. The encoder $E_\theta$ takes the initial condition $x_0 \in \mathbb{R}^n$ and maps it into the latent space $z_0 \in \mathbb{R}^m$. The neural ODE adopted is an autonomous system that does not explicitly dependent on independent variables i.e $t$. $z_t$ can be obtained by integrating the neural ODE from $t = 0$ to $t$. The Decoder $D_\theta$ decode the latent vector $z_t$ alongside the initial latent vector and the back to the abundance space $x_{\rm t, pred}$.  }
\vspace{-0.8cm}
\label{fig:schem}
\end{center}
\end{figure}

\section{Model}
\label{sec:model}
Our proposed model consist of three separate neural networks, the Encoder $E_\theta$, Decoder $D_\theta$, and the neural ODE function $f_\theta$ and a schematic of the architecture is shown in Figure \ref{fig:schem}.
The overall model can be specified with the equations below. 
\begin{align}
    \mathbf{z}_0 &= E_\theta (\mathbf{ \tilde{x}_0 }), \quad \mathbf{x} \in \mathbb{R}^n \nonumber \\
    \frac{d \mathbf{z}}{dt} &= g_\theta(t, \mathbf{z}), \quad \mathbf{z} \in \mathbb{R}^m \\
    \mathbf{z}_t  &= \mathbf{z}_0 + \int_0^t g_\theta( \mathbf{z}_{t'}) dt' \nonumber\\
    \mathbf{\tilde{x}}_{\rm t, pred} &= \tilde{D}_\theta (\mathbf{z}_t) \nonumber \\
    \mathbf{x}_{\rm t, pred} &= D_\theta (\mathbf{z}_t, z_0, x_0 ) \nonumber
    \label{eq:model}
\end{align}
The encoder $E_\theta$ is a learnable function that takes the $\log$-normalized state vector as input $\tilde{\mathbf{x}} \in \mathbb{R}^{n}$, and maps it to the latent space $z \in \mathbb{R}^m$. 
In our experiments, the dimension of the state vector $n$ and the latent vectors $m$ are 10 and 3 respectively.
Hereafter $\tilde{\mathbf{x}}$ corresponds to $\log$-normalized state vector and $\mathbf{x}$ refers to the unnormalized state vector. 
The RHS function $g_\theta$ defines the dynamics of the ODE in the latent space. 
The encoder $E_\theta: \mathbb{R}^{\rm n} \rightarrow \mathbb{R}^{m} $ and the ODE function $g_\theta: \mathbb{R}^{\rm m} \rightarrow \mathbb{R}^{m} $ are parametrized with a multi-layer perceptron (MLP) with 4 layers and 32 hidden units with the \texttt{ELU} activation function.
In order to limit the potentially arbitrarily large $\frac{dz}{dt}$ learnt by the neural ODE network, we have further applied an additional $\texttt{tanh}$ activation with a learnable scale factor $\mathbf{s}_z \in \mathbb{R}^m$ to stabilize the flow field learned by the neural network, i.e. $ \frac{d \mathbf{z} }{dt} = \rm \mathbf{s}_z \circ ~ tanh( MLP_\theta( \mathbf{z}) /~ \mathbf{s}_z)  $. 
By restricting the limit of the $\frac{dz}{dt}$ to between $[-\mathbf{s}_z, \mathbf{s}_z]$, it is empirically  effective at stabilizing training in our experiments. 

As for the decoder $D_\theta$, we have experimented with two different architectures. One of them is a MLP  $\tilde{D}_\theta: \mathbb{R}^{\rm m} \rightarrow \mathbb{R}^{n} $ that takes the latent vector $z$ as input and output the abundance in $\log$-normalized space $\tilde{\mathbf{x}} \in \mathbb{R}^n$. We termed it the ``vanilla decoder". It is also parameterized with a 4-layer MLP with 32 hidden units with \texttt{ELU} activation function.
This setup is similar to the one proposed in \citet{grassi2021}, except that in our work the latent network dynamics is replaced with a more versatile MLP $g_\theta$. We have also device a new type of decoder that incorporates the initial condition of the ODE $D_\theta: \mathbb{R}^{\rm 2m+n} \rightarrow \mathbb{R}^{n} $, $\mathbf{x}_{t,\rm pred} = \mathbf{x}_0 \circ ~ \exp{( \rm \mathbf{s} ~ \circ ~ MLP_\theta(z_t, z_0, \tilde{x}_0))}$. Here $ \circ$ corresponds to elementwise multiplication.

 Instead of directly learning the mapping from the latent space $\mathbf{z}$ to the abundance space $\mathbf{x}$, we have the decoder predict the $\log$-variation of the abundance with respect to its initial abundance. This kind of parametrization also guarantees the output from the decoder is always positive definite. $\mathbf{s} \in \mathbb{R_+}^n$ here corresponds to a learnable scale factor that accounts for the scale over which the abundance varied across the timescale of interest. For comparison, we have also used a vanilla encoder-decoder combination with a single linear layer that mimics a latent space generated from linear projection.
 
One of the measure of the dynamical timescale is the free-fall timescale $t_{\rm ff}$ and it can be estimated directly by $t_{\rm ff} \approx \frac{1}{\sqrt{G\rho}}$, where $G$ is the gravitational constant, and $\rho$ is the density of the cell. In order to accommodate a wide range of dynamical timescale of interest, time axis for each of the trajectory $t$ are normalized by the respective freefall timescale given by the density implied from the initial condition $\mathbf{x}_0$.

The loss function $\mathcal{L}$ for each given trajectory is specified in terms of the true solution $\mathbf{x}_t$ and the predicted solution $\mathbf{x}_{\rm t, pred}$ as below:
\begin{align*}
    \mathcal{L}_{\rm path}(\mathbf{x}, \mathbf{x}_{\rm pred}) =&  \sum_{t} |\log (\mathbf{x}_{\rm t}) -  \log(\mathbf{x}_{\rm t, pred}) |, \\
    \mathcal{L}_{\rm recon}(\mathbf{x}) =&  \sum_{t} |\log{(\mathbf{x}_t)} - \log{ D_\theta ( E_\theta (\mathbf{x}_t), \mathbf{z}_0, \mathbf{x}_0 )} |, \\
    \mathcal{L}_{\rm conserv}(\mathbf{x}) =&  \mathcal{L}_{\rm conserv, H}(\mathbf{x}) \\ 
    &+ \mathcal{L}_{\rm conserv, He}(\mathbf{x}) + \mathcal{L}_{\rm conserv, e}(\mathbf{x}) \\
    \mathcal{L} =\mathcal{L}_{\rm path} &+ \lambda_1 \mathcal{L}_{\rm recon}(x) + \lambda_2 \mathcal{L}_{\rm conserv} (x_{\rm pred}) 
\end{align*}
$\mathcal{L}_{\rm path}$ corresponds to the absolute $\log$-difference of the predicted values and the actual trajectory, and $\mathcal{L}_{\rm recon}$ enforces that the initial condition is consistently encoded in the latent space. This is often termed the reconstruction loss in the context of an autoencoder, where the objective of an autoencoder is to reconstruct the state vector $x$. The total mass density in hydrogen $\rm H$ and helium $\rm He$ should be kept the same as the time progress. The net charge of each trajectory should also stay zero in the reconstructed solution. $\mathcal{L}_{\rm conserv}$ encourages the system to penalize solutions that do not obey the law of conservation. $\lambda_1$, $\lambda_2$ are both set to unity in the rest of our experiments. The loss function defined above is specified for our proposed initial-condition guided autoencoder. This is defined similarly for the $\log$-normalized outputs from the plain decoder.

These three neural networks are optimized jointly to minimize the loss function $\mathcal{L}$. 
We have made used of the \texttt{torchdiffeq} package to perform ODE integration, and backpropagation through the ODE solution using the memory-efficient adjoint method \cite{chen2018neuralode}. We refer our readers to Appendix \ref{app:train} for the details on the implementation and training. The code are made available on Github\footnote{https://github.com/hisunnytang/neuralODE}.








\section{Results}

\subsection{Trajectory Prediction and Accuracy}
\begin{table}
\centering
\begin{tabular}{c|c|c|c|c}
      $\rm 10^{-3}$&  1-layer & Vanilla & Our Model\\
     \hline
     $\mathcal{L}_{\rm path}$  & 296 & 74.3 & $\mathbf{6.10}$  \\
     $\mathcal{L}_{\rm recon}$ & 288 & 84.2 & $\mathbf{5.35}$ \\
     $ \mathcal{L}_{\rm conserv}$ & 2.04 & 11.8  & $\mathbf{0.10}$  \\
     Mean Pct. Err. (\%) & 36.0   & 9.00  &  $\mathbf{0.97}$ \\
     \hline
\end{tabular}
\caption{Error Metrics for three of the models experimented. Each of the loss term is presented in $\rm 10^{-3}$. The last row corresponds to the mean percentage error evaluated in the physical space.}
\vspace{-0.5cm}
\label{tab:accuracy}
\end{table}    

\begin{figure}[!ht]
    \centering
    \includegraphics[width=\linewidth]{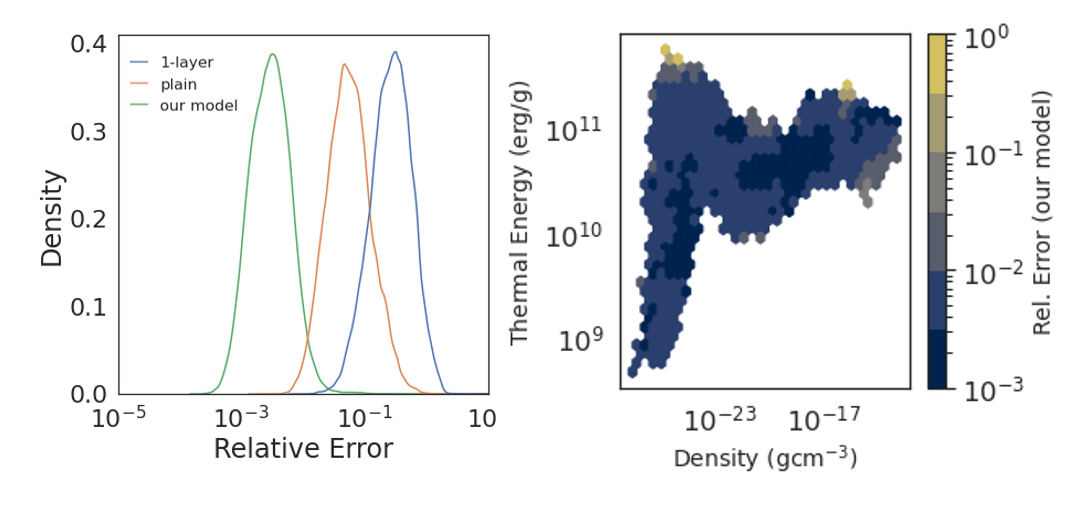}
    \vspace{-0.5cm}
    \caption{Left-Panel: Relative Error Distribution calculated in the physical space (not log-normalized space) for the three different models experimented. Right-Panel: Relative error distribution in the density-thermal energy phase space our initial value guided autoencoder. It on average shows an relative error of less than $1\%$ across the phase space, except for regions with high temperatures.}
    \label{fig:error_dist}
\end{figure}

The summary statistics and the relative error distribution of the three models on the test-set data is tabulated and reported in Table \ref{tab:accuracy} and in Figure \ref{fig:error_dist}. The absolute log error decreases with increasing model complexity. The one-layer model which mimics a simple linear projection performs the worst out of these models. This indicates a simple linear projection mapping is incapable of finding a good latent space for proper reconstruction. This also motivates and justifies the use of a more complicated autoencoder that is able to capture the non-linear interactions between the chemical species. Our plain-autoencoder architecture, which shares the most resemblance with the one proposed by \citet{grassi2021}, shows on average a  $\rm 10\%$ relative error. Our proposed initial-value guided decoder architecture achieves the best performance out of all three models considered with a relative error of $\rm 1\%$. On the right panel of Figure \ref{fig:error_dist}, we showed the distribution of relative error across the density-thermal-energy phase space. It shows on average an relative error of less than $\rm 1\%$ across the phase space, except for regions with high temperatures and high density.

\begin{figure*}[!ht]
    \centering
    \includegraphics[width=\textwidth]{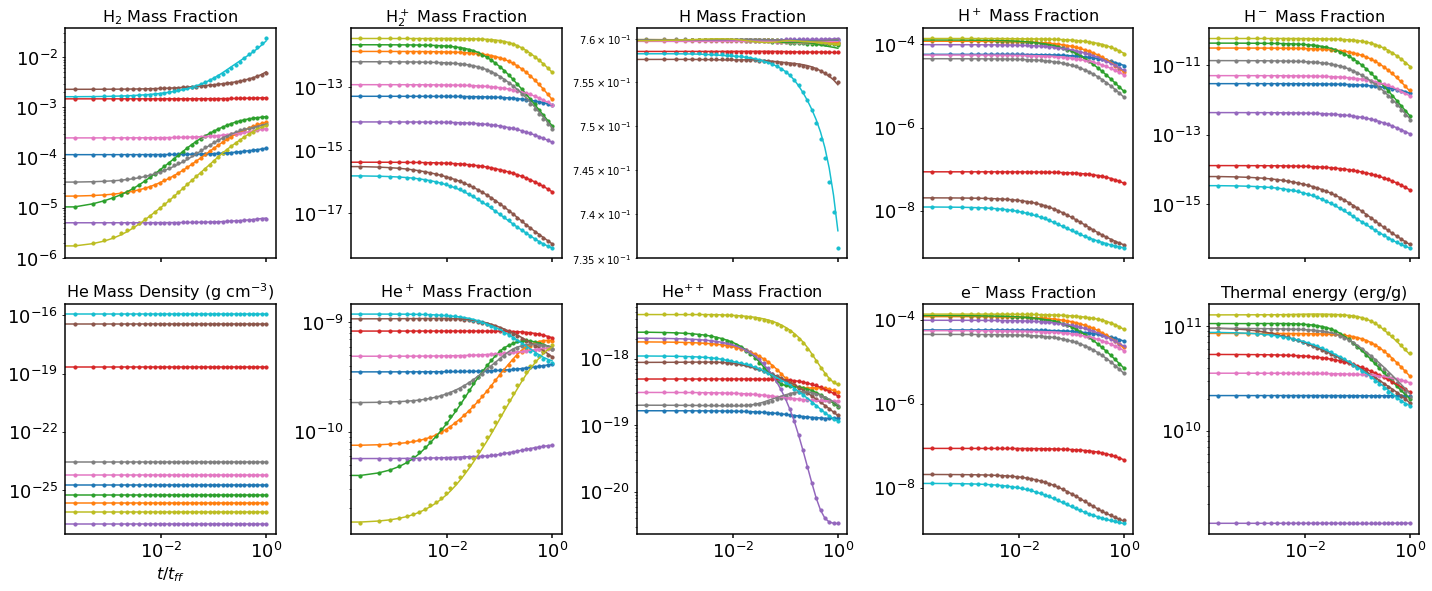}
    \vspace{-0.5cm}
    \caption{Sample Trajectories from various initial conditions with our proposed initial value guided autoencoder. The time axis are all normalized to the freefall timescale. The dots represent the ground-truth solution generated from \texttt{Dengo} solver with a relative tolerance of $10^{-6}$. The lines show the respective trajectories from our latent ODE integration. Different initial conditions are highlighted in different colors. Except for \texttt{He} and thermal energy, the rest of the species of interest are normalized by the total density and expressed in mass fractions. Our model is capable of handling initial conditions that spans $10$ orders of magnitude in density (can be inferred from the He Mass Density Panel), more than $2$ orders of magnitude in thermal energy across various freefall timescales defined by the density. Sample trajectories from our 1-layer autoencoder, and vanilla autoencoder is also shown in Appendix \ref{app:traj}. }
    \label{fig:traj}
\end{figure*}

As a demonstration, various initial conditions are drawn from the test set, and are evolved with our trained models. The true trajectories are shown alongside with the predicted trajectories in Figure \ref{fig:traj}. Despite the wide range of order of magnitude involved across not only density and temperature, but also the species abundances, our model is still capable of recovering the morphology of the correct trajectories up to one freefall timescale. We refer our readers to Appendix \ref{app:traj} for the trajectories from the rest of the two models.

\subsection{Performance Comparison}
In this section, we look into the performance of our proposed model and \texttt{Grackle}. The learnt neural ODE, Encoder, Decoder are exported to \texttt{TorchScript} from the \texttt{PyTorch} \cite{pytorch} model checkpoint. The resulting \texttt{TorchScript} model can be run independently from \texttt{Python} as a standalone \texttt{C++} program in a production environment. Similar to the above described procedures, the abundances are transformed into the latent space and from latent space to abundances spaces with the Encoder and Decoder model respectively. The numerical integration is performed with the \texttt{CVode} \cite{hindmarsh2005sundials} with the neural ODE function learnt in the latent space $z$. Since the neural ODE model is fully differentiable, the respective Jacobian $\frac{\partial g_\theta}{\partial z}$ can be constructed with almost no cost with the the \texttt{autograd} function available to the \texttt{PyTorch} model. The table below shows a comparison between our various proposed models. 

\begin{table}[h!]
\centering
\begin{tabular}{c|c|c}
     Per-cell runtime ($10^{-6}\rm s$) &  0.1 $t_{\rm ff}$ &  $t_{\rm ff}$ \\
     \hline
     \texttt{Grackle} & $39.2 \pm 2.8$ & $41.6\pm3.1$ \\
     \texttt{TorchScipt+CVODE} & $\bf{3.83 \pm 0.05}$ & $\bf{6.32 \pm 0.10}$ \\
     \hline
\end{tabular}
\caption{Runtime Comparison in production environment among the \texttt{Grackle} and Our Model.}
\label{tab:runtime}
\end{table}    

The experiments are performed with the test set data with  on \texttt{Intel(R) Xeon(R) CPU E5-2650 v3 @ 2.30GHz} with 40-threads \texttt{OpenMP} acceleration. The \texttt{TorchScript} model are deployed in parallel with batch size of 2048. The results are shown and tabulated in Table \ref{tab:runtime}. Our proposed model with modest parallelization has achieved an almost tenfold speed up compared to the widely adopted primordial chemistry solver \texttt{Grackle} on this particular 9-species model.



\section{Discussion and Future Work}

We introduce our new architecture for reducing chemical network in a data-driven way and show that our proposed architecture is capable of recovering the temporal trajectories faithfully across a wide-range of initial conditions in a deployment setting with one-tenth of the runtime compared to a commonly used astro-chemistry library. We note that when comparing the performance between our solvers and \texttt{Grackle}, our models are deployed on CPU only. By extending the current \texttt{Torchscript} model and its interface with \texttt{CVODE} to GPU \cite{balos2021enabling}, we should expect to see a further speedup in terms of the runtime. The model can be further improved by adding an additional terms in the loss function that penalize the stiffness of the latent space dynamics equation. Such regularization can implicitly enforce the learnt vector field $g_\theta$ in the latent space to be smoother and render it easier to integrate.


\bibliography{main}

\begin{thebibliography}{22}
\providecommand{\natexlab}[1]{#1}
\providecommand{\url}[1]{\texttt{#1}}
\expandafter\ifx\csname urlstyle\endcsname\relax
  \providecommand{\doi}[1]{doi: #1}\else
  \providecommand{\doi}{doi: \begingroup \urlstyle{rm}\Url}\fi

\bibitem[Baldi \& Hornik(1989)Baldi and Hornik]{BALDI198953}
Baldi, P. and Hornik, K.
\newblock Neural networks and principal component analysis: Learning from
  examples without local minima.
\newblock \emph{Neural Networks}, 2\penalty0 (1):\penalty0 53--58, 1989.
\newblock ISSN 0893-6080.
\newblock \doi{https://doi.org/10.1016/0893-6080(89)90014-2}.
\newblock URL
  \url{https://www.sciencedirect.com/science/article/pii/0893608089900142}.

\bibitem[Balos et~al.(2021)Balos, Gardner, Woodward, and
  Reynolds]{balos2021enabling}
Balos, C.~J., Gardner, D.~J., Woodward, C.~S., and Reynolds, D.~R.
\newblock {Enabling GPU accelerated computing in the SUNDIALS time integration
  library}.
\newblock \emph{Parallel Computing}, 108:\penalty0 102836, 2021.

\bibitem[Brunton et~al.(2020)Brunton, Noack, and Koumoutsakos]{ml_fluid}
Brunton, S.~L., Noack, B.~R., and Koumoutsakos, P.
\newblock Machine learning for fluid mechanics.
\newblock \emph{Annual Review of Fluid Mechanics}, 52\penalty0 (1):\penalty0
  477--508, 2020.
\newblock \doi{10.1146/annurev-fluid-010719-060214}.
\newblock URL \url{https://doi.org/10.1146/annurev-fluid-010719-060214}.

\bibitem[Bryan et~al.(2014)Bryan, Norman, O{\textquotesingle}Shea, Abel, Wise,
  Turk, Reynolds, Collins, Wang, Skillman, Smith, Harkness, Bordner, hoon Kim,
  Kuhlen, Xu, Goldbaum, Hummels, Kritsuk, Tasker, Skory, Simpson, Hahn, Oishi,
  So, Zhao, Cen, and and]{Bryan_2014}
Bryan, G.~L., Norman, M.~L., O{\textquotesingle}Shea, B.~W., Abel, T., Wise,
  J.~H., Turk, M.~J., Reynolds, D.~R., Collins, D.~C., Wang, P., Skillman,
  S.~W., Smith, B., Harkness, R.~P., Bordner, J., hoon Kim, J., Kuhlen, M., Xu,
  H., Goldbaum, N., Hummels, C., Kritsuk, A.~G., Tasker, E., Skory, S.,
  Simpson, C.~M., Hahn, O., Oishi, J.~S., So, G.~C., Zhao, F., Cen, R., and
  and, Y.~L.
\newblock {ENZO}: {AN} {ADAPTIVE} {MESH} {REFINEMENT} {CODE} {FOR}
  {ASTROPHYSICS}.
\newblock \emph{The Astrophysical Journal Supplement Series}, 211\penalty0
  (2):\penalty0 19, mar 2014.
\newblock \doi{10.1088/0067-0049/211/2/19}.
\newblock URL \url{https://doi.org/10.1088%2F0067-0049%2F211%2F2%2F19}.

\bibitem[Champion et~al.(2019)Champion, Lusch, Kutz, and Brunton]{champion2019}
Champion, K., Lusch, B., Kutz, J.~N., and Brunton, S.~L.
\newblock Data-driven discovery of coordinates and governing equations, 2019.
\newblock URL \url{https://arxiv.org/abs/1904.02107}.

\bibitem[Chen et~al.(2018)Chen, Rubanova, Bettencourt, and
  Duvenaud]{chen2018neuralode}
Chen, R. T.~Q., Rubanova, Y., Bettencourt, J., and Duvenaud, D.
\newblock Neural ordinary differential equations.
\newblock \emph{Advances in Neural Information Processing Systems}, 2018.

\bibitem[{Gong} et~al.(2017){Gong}, {Ostriker}, and {Wolfire}]{Gong_2017}
{Gong}, M., {Ostriker}, E.~C., and {Wolfire}, M.~G.
\newblock {A Simple and Accurate Network for Hydrogen and Carbon Chemistry in
  the Interstellar Medium}.
\newblock \emph{ApJ}, 843\penalty0 (1):\penalty0 38, July 2017.
\newblock \doi{10.3847/1538-4357/aa7561}.

\bibitem[Goodfellow et~al.(2016)Goodfellow, Bengio, and
  Courville]{GoodBengCour16}
Goodfellow, I.~J., Bengio, Y., and Courville, A.
\newblock \emph{Deep Learning}.
\newblock MIT Press, Cambridge, MA, USA, 2016.
\newblock \url{http://www.deeplearningbook.org}.

\bibitem[Grassi et~al.(2021)Grassi, Nauman, Ramsey, Bovino, Picogna, and
  Ercolano]{grassi2021}
Grassi, T., Nauman, F., Ramsey, J.~P., Bovino, S., Picogna, G., and Ercolano,
  B.
\newblock Reducing the complexity of chemical networks via interpretable
  autoencoders, 2021.
\newblock URL \url{https://arxiv.org/abs/2104.09516}.

\bibitem[{Hahn} \& {Abel}(2011){Hahn} and {Abel}]{MUSIC}
{Hahn}, O. and {Abel}, T.
\newblock {Multi-scale initial conditions for cosmological simulations}.
\newblock \emph{MNRAS}, 415\penalty0 (3):\penalty0 2101--2121, August 2011.
\newblock \doi{10.1111/j.1365-2966.2011.18820.x}.

\bibitem[Hindmarsh et~al.(2005)Hindmarsh, Brown, Grant, Lee, Serban, Shumaker,
  and Woodward]{hindmarsh2005sundials}
Hindmarsh, A.~C., Brown, P.~N., Grant, K.~E., Lee, S.~L., Serban, R., Shumaker,
  D.~E., and Woodward, C.~S.
\newblock {SUNDIALS}: Suite of nonlinear and differential/algebraic equation
  solvers.
\newblock \emph{ACM Transactions on Mathematical Software (TOMS)}, 31\penalty0
  (3):\penalty0 363--396, 2005.

\bibitem[{Kamp} et~al.(2017){Kamp}, {Thi}, {Woitke}, {Rab}, {Bouma}, and
  {M{\'e}nard}]{Kamp_2017}
{Kamp}, I., {Thi}, W.~F., {Woitke}, P., {Rab}, C., {Bouma}, S., and
  {M{\'e}nard}, F.
\newblock {Consistent dust and gas models for protoplanetary disks. II.
  Chemical networks and rates}.
\newblock \emph{AAP}, 607:\penalty0 A41, November 2017.
\newblock \doi{10.1051/0004-6361/201730388}.

\bibitem[{Omukai}(2000)]{Omukai_2000}
{Omukai}, K.
\newblock {Protostellar Collapse with Various Metallicities}.
\newblock \emph{ApJ}, 534\penalty0 (2):\penalty0 809--824, May 2000.
\newblock \doi{10.1086/308776}.

\bibitem[{Omukai} et~al.(2005){Omukai}, {Tsuribe}, {Schneider}, and
  {Ferrara}]{Omukai_2005}
{Omukai}, K., {Tsuribe}, T., {Schneider}, R., and {Ferrara}, A.
\newblock {Thermal and Fragmentation Properties of Star-forming Clouds in
  Low-Metallicity Environments}.
\newblock \emph{ApJ}, 626\penalty0 (2):\penalty0 627--643, June 2005.
\newblock \doi{10.1086/429955}.

\bibitem[Paszke et~al.(2019)Paszke, Gross, Massa, Lerer, Bradbury, Chanan,
  Killeen, Lin, Gimelshein, Antiga, Desmaison, Kopf, Yang, DeVito, Raison,
  Tejani, Chilamkurthy, Steiner, Fang, Bai, and Chintala]{pytorch}
Paszke, A., Gross, S., Massa, F., Lerer, A., Bradbury, J., Chanan, G., Killeen,
  T., Lin, Z., Gimelshein, N., Antiga, L., Desmaison, A., Kopf, A., Yang, E.,
  DeVito, Z., Raison, M., Tejani, A., Chilamkurthy, S., Steiner, B., Fang, L.,
  Bai, J., and Chintala, S.
\newblock Pytorch: An imperative style, high-performance deep learning library.
\newblock In \emph{Advances in Neural Information Processing Systems 32}, pp.\
  8024--8035. Curran Associates, Inc., 2019.
\newblock URL
  \url{http://papers.neurips.cc/paper/9015-pytorch-an-imperative-style-high-performance-deep-learning-library.pdf}.

\bibitem[{Richings} et~al.(2014{\natexlab{a}}){Richings}, {Schaye}, and
  {Oppenheimer}]{2014MNRAS.440.3349R}
{Richings}, A.~J., {Schaye}, J., and {Oppenheimer}, B.~D.
\newblock {Non-equilibrium chemistry and cooling in the diffuse interstellar
  medium - I. Optically thin regime}.
\newblock \emph{MNRAS}, 440\penalty0 (4):\penalty0 3349--3369, June
  2014{\natexlab{a}}.
\newblock \doi{10.1093/mnras/stu525}.

\bibitem[{Richings} et~al.(2014{\natexlab{b}}){Richings}, {Schaye}, and
  {Oppenheimer}]{2014MNRAS.442.2780R}
{Richings}, A.~J., {Schaye}, J., and {Oppenheimer}, B.~D.
\newblock {Non-equilibrium chemistry and cooling in the diffuse interstellar
  medium - II. Shielded gas}.
\newblock \emph{MNRAS}, 442\penalty0 (3):\penalty0 2780--2796, August
  2014{\natexlab{b}}.
\newblock \doi{10.1093/mnras/stu1046}.

\bibitem[{Smith} et~al.(2017){Smith}, {Bryan}, {Glover}, {Goldbaum}, {Turk},
  {Regan}, {Wise}, {Schive}, {Abel}, {Emerick}, {O'Shea}, {Anninos}, {Hummels},
  and {Khochfar}]{Grackle}
{Smith}, B.~D., {Bryan}, G.~L., {Glover}, S.~C.~O., {Goldbaum}, N.~J., {Turk},
  M.~J., {Regan}, J., {Wise}, J.~H., {Schive}, H.-Y., {Abel}, T., {Emerick},
  A., {O'Shea}, B.~W., {Anninos}, P., {Hummels}, C.~B., and {Khochfar}, S.
\newblock {GRACKLE: a chemistry and cooling library for astrophysics}.
\newblock \emph{MNRAS}, 466:\penalty0 2217--2234, April 2017.
\newblock \doi{10.1093/mnras/stw3291}.

\bibitem[{Tang} \& {Turk}(In Prep){Tang} and {Turk}]{dengo}
{Tang}, K.~S. and {Turk}, M.~J.
\newblock Dengo: An engineer for chemistry solvers, In Prep.

\bibitem[{Turk} et~al.(2011){Turk}, {Smith}, {Oishi}, {Skory}, {Skillman},
  {Abel}, and {Norman}]{yt_turk_2011}
{Turk}, M.~J., {Smith}, B.~D., {Oishi}, J.~S., {Skory}, S., {Skillman}, S.~W.,
  {Abel}, T., and {Norman}, M.~L.
\newblock {yt: A Multi-code Analysis Toolkit for Astrophysical Simulation
  Data}.
\newblock \emph{ApjS}, 192\penalty0 (1):\penalty0 9, January 2011.
\newblock \doi{10.1088/0067-0049/192/1/9}.

\bibitem[Vinuesa \& Brunton(2021)Vinuesa and Brunton]{VinuesaBrunton2021}
Vinuesa, R. and Brunton, S.~L.
\newblock The potential of machine learning to enhance computational fluid
  dynamics, 2021.
\newblock URL \url{https://arxiv.org/abs/2110.02085}.

\bibitem[Vlachas et~al.(2022)Vlachas, Arampatzis, Uhler, and
  Koumoutsakos]{Vlachas2022}
Vlachas, P.~R., Arampatzis, G., Uhler, C., and Koumoutsakos, P.
\newblock Multiscale simulations of complex systems by learning their effective
  dynamics.
\newblock \emph{Nature Machine Intelligence}, 4\penalty0 (4):\penalty0
  359--366, Apr 2022.
\newblock ISSN 2522-5839.
\newblock \doi{10.1038/s42256-022-00464-w}.
\newblock URL \url{https://doi.org/10.1038/s42256-022-00464-w}.

\end{thebibliography}
\bibliographystyle{icml2022}

\newpage
\appendix
\onecolumn
\section{Cosmological Simulation and Data Sampling}
\label{app:sims}

The simulation is initialized at $z = 50$ in a $0.3 ~ \mathrm{Mpc} /h$ co-moving box with \texttt{MUSIC} \citep{MUSIC} initial conditions generator. We have first run a dark matter only simulation to identify the isolated halos of interest with mass ranges between $10^5 - 10^6 M_{\odot}$. These initial conditions are re-centered and generated again based on the location of the most massive halo identified above. Cells are flagged for refinement if the local Jeans length is not resolved by 64 cells. Evolution timescale is also limited to a tenth of the thermal timescale. A snapshot at $z = 18$ is taken as our initial state vector for the generation of the trajectory dataset.

\begin{figure}[!ht]
    \centering
    \includegraphics[width=0.8\linewidth]{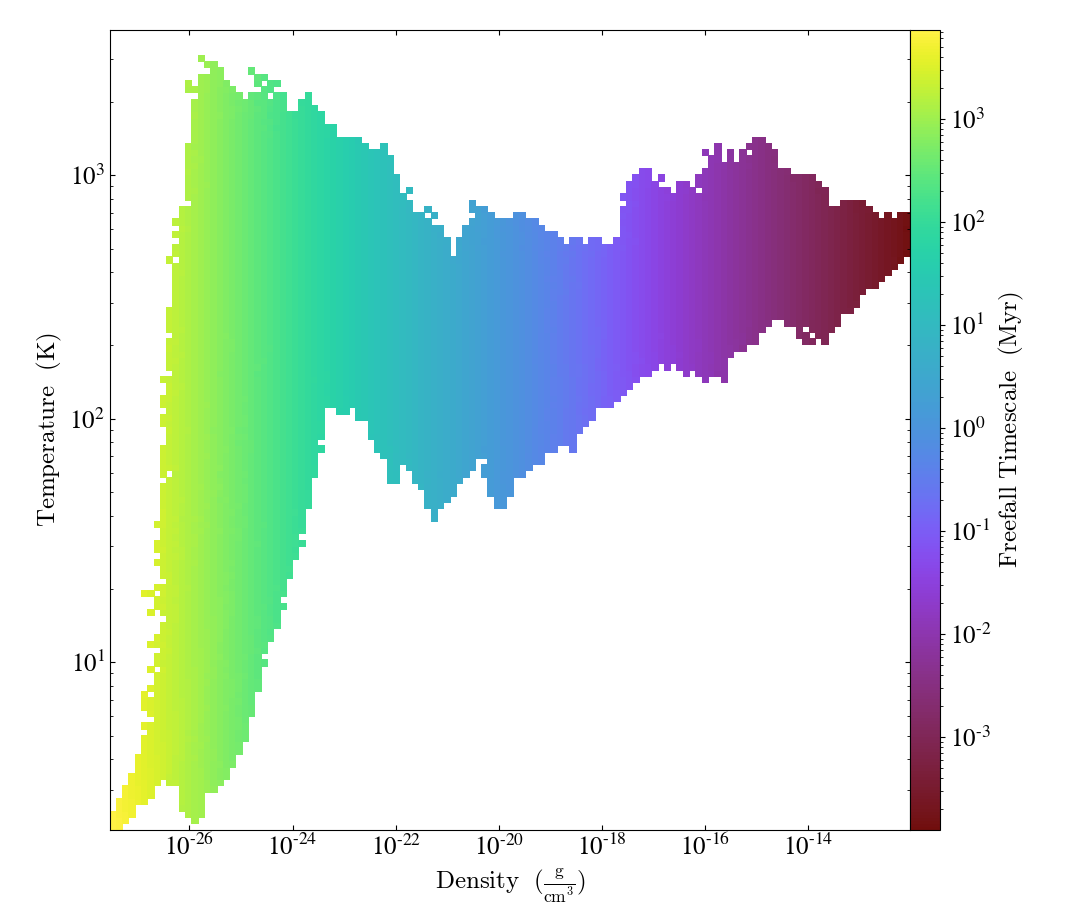}
    \caption{The phase distribution of the density and temperature of the initial conditions taken from the simulation snapshot. The colorbar demonstrates the characteristic freefall timescale which spans from $\rm 10^{-3} - 10^{3} Myr$.}
    \label{fig:phaseplot}
\end{figure}

For any user-specified chemistry network, \texttt{Dengo} \cite{dengo} can generate the corresponding RHS function ($f$), the Jacobian function ($\frac{\partial f}{\partial y}$) and the interface to solve the system of ODE with integrator \texttt{Sundials CVODE} \cite{hindmarsh2005sundials}. The availability of the Jacobian allows for a noticeable speed up in the integration process particular for stiff ODE system. The relative tolerance of \texttt{Dengo} is set to $10^{-6}$ when the trajectory is generated. The output timesteps are logarithimcally spaced. There are a total of 3068798 number of grid cells. As a pilot study, we limit our dataset to a subset that is representative of the physical conditions in the simulations. The dataset is sub-sampled on a $64\times 64\times 64$ grid of density, temperature and $\rm H_2$ mass fraction, where in each unique bin, 5 samples are select. This leaves us with 114401 cells. It is split into train-validation-test set with a ratio of $\rm 0.8: 0.1: 0.1$ The data analysis and data extraction are performed with $\texttt{yt}$ \cite{yt_turk_2011}.

\section{Model and Training Details}
\label{app:train}
\begin{figure}[!ht]
    \centering
    \includegraphics[width=0.7\linewidth]{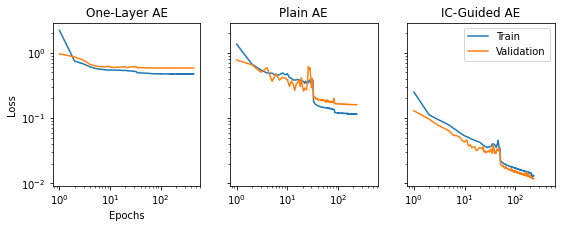}
    \caption{Training and validation loss for the three models experimented in this work.}
    \label{fig:loss}
\end{figure}

The models outlined in Section \ref{sec:model} are all implemented in \texttt{PyTorch} \cite{pytorch}. The models are trained with \texttt{Adam} optimizer with an initial learning rate of $\rm 10^{-3}$\texttt{ReduceLROnPlateau} learning rate scheduler with default parameters and a minimum learning rate of $\rm 10^{-6}$. The training is stopped early with a patience of 5 epochs. The training and validation curve for the various models are shown in Figure \ref{fig:loss}.
\newpage
\section{Trajectories of 1-layer, and Plain Autoencoder architecture}
\label{app:traj}

Similar to Fig. \ref{fig:traj}, the trajectories for various initial conditions are shown for the 1-layer model, and the plain autoencoder model. Our proposed model shows a 

\begin{figure}[!ht]
    \centering
    \includegraphics[width=\linewidth]{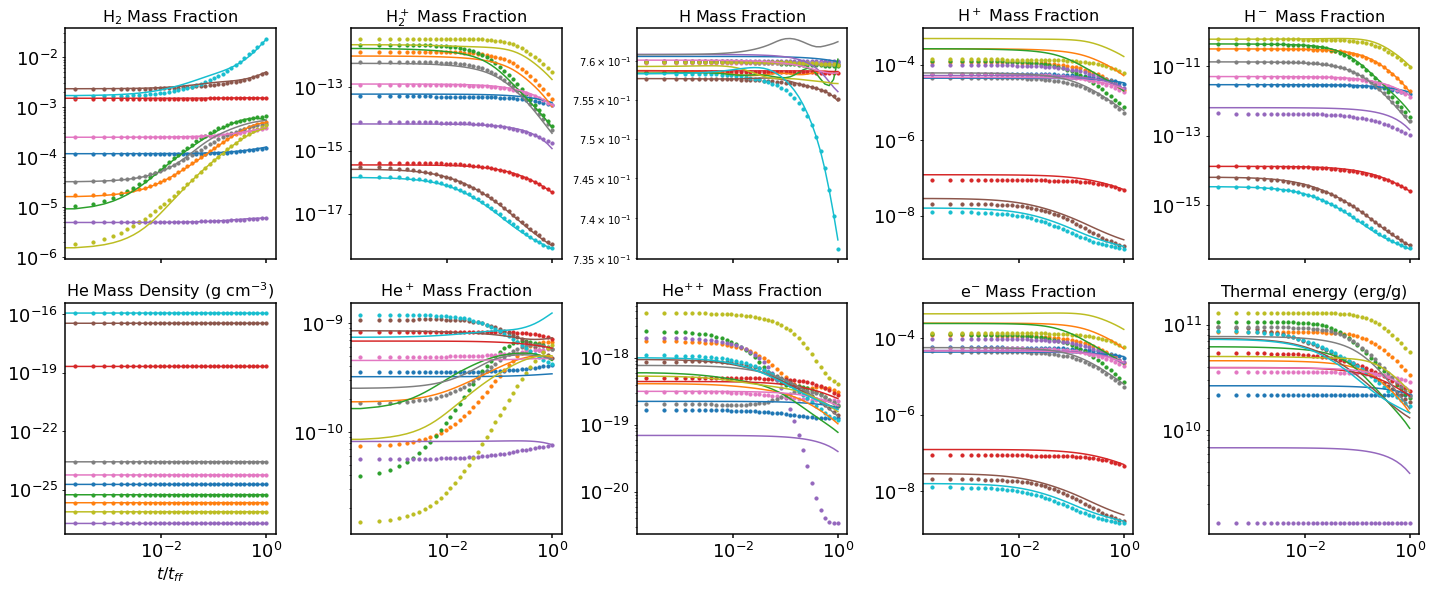}
    \caption{Trajectory Predictions from our 1-layer autoencoder model}
    \label{fig:1layer_traj}
\end{figure}

\begin{figure}[!ht]
    \centering
    \includegraphics[width=\linewidth]{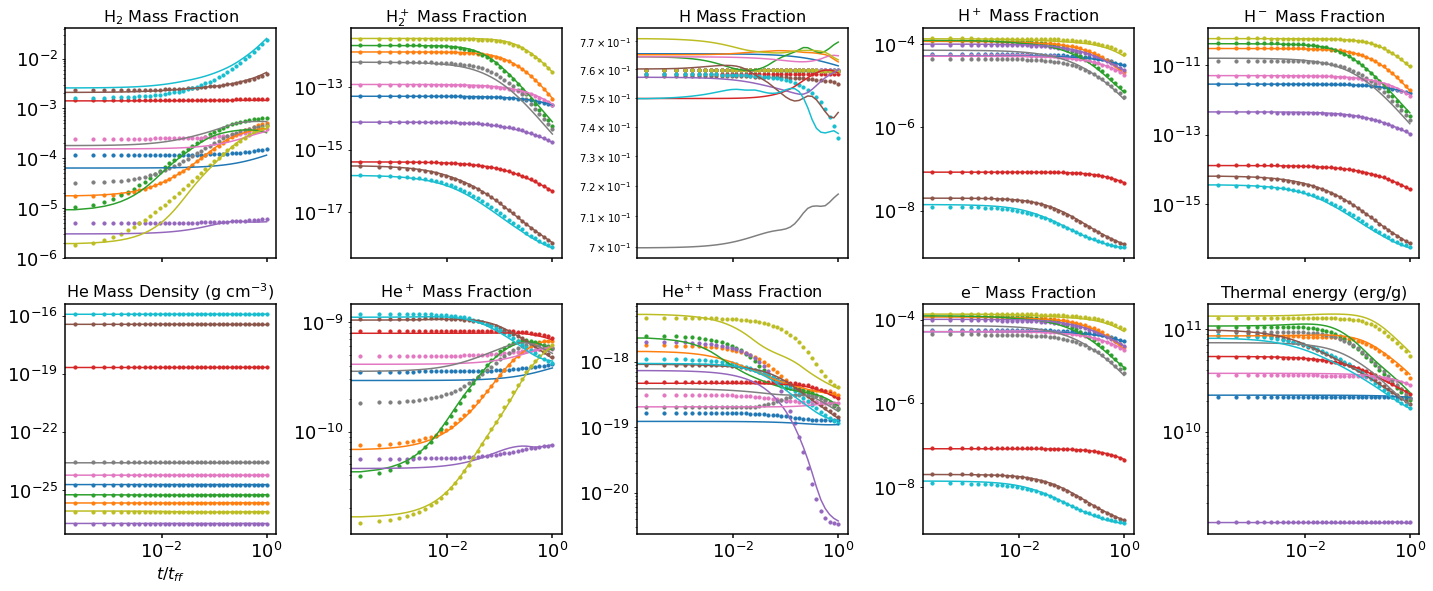}
    \caption{Trajectory Predictions from our Plain autoencoder model}
    \label{fig:plain_traj}
\end{figure}

\newpage
\section{Error Distribution of the 1-layer, Plain Autoencoder and our model}
\label{app:err}
\begin{figure}[!ht]
    \centering
    \includegraphics[width=\linewidth]{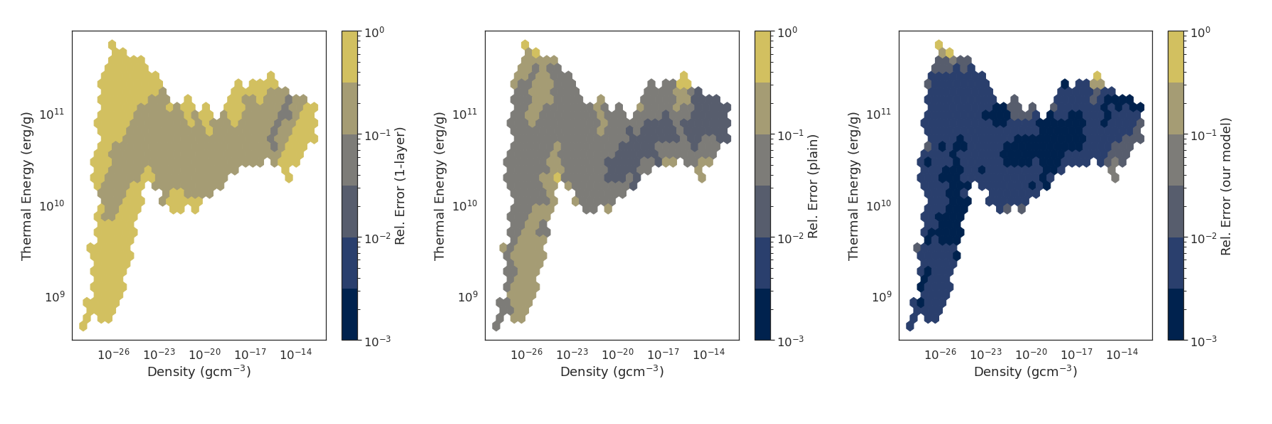}
    \caption{The error distribution of the three models presented in density-temperature phase space}
    \label{fig:my_label}
\end{figure}

\end{document}